\documentclass[reprint,superscriptaddress,amsmath,amssymb,aps,]{revtex4-2}
\usepackage{amssymb}
\usepackage[font=small,labelfont=bf,justification=justified]{caption}
\usepackage{amsmath}
\usepackage{amsfonts}
\usepackage{graphicx}
\usepackage{hyperref}
\usepackage{times}
\usepackage{array}
\usepackage{graphicx}
\usepackage{mathptmx}
\usepackage{algorithmic}
\usepackage{textcomp}
\usepackage{algorithm}
\usepackage{url}
\usepackage{physics}
\usepackage{verbatim}
\usepackage{braket}
\usepackage{tabularx}
\usepackage{amsthm}
\usepackage{mathrsfs}
\usepackage{xcolor}
\usepackage{gensymb} 
\usepackage{caption}
\usepackage{subcaption}

\begin{document}

\title{Investigating a Device Independence Quantum Random Number Generation}

\begin{abstract}
    Quantum random number generation (QRNG) is a resource that is a necessity in the field of cryptography. However, its certification has been challenging. In this article, we certify randomness with the aid of quantum entanglement in a device independent setting, where we choose two-photon interference for source characterisation. The CHSH inequality violation and quantum state tomography are used as independent checks on the measurement devices.  These measures ensure the unpredictability of quantum random number generation. This work can be easily extended to faster randomness expansion protocols.
\end{abstract}

\author{Vardaan Mongia}
\affiliation{Quantum Science and Technology Laboratory, Physical Research Laboratory, Ahmedabad, 380009, India}
\affiliation{Dept. of Physics, Indian Institute of Technology Gandhinagar, Gujarat, 382355, India}

\author{Abhishek Kumar}
\affiliation{Space Weather Laboratory, Physical Research Laboratory, Ahmedabad, 380009, India}

\author{Shashi Prabhakar}
\affiliation{Quantum Science and Technology Laboratory, Physical Research Laboratory, Ahmedabad, 380009, India}

\author{Anindya Banerji}
\affiliation{Center for Quantum Technologies, National University of Singapore, Singapore 117543}

\author{R.P. Singh}
\affiliation{Quantum Science and Technology Laboratory, Physical Research Laboratory, Ahmedabad, 380009, India}

% \affil[1]{Quantum Science and Technology Laboratory, Physical Research Laboratory, Ahmedabad, 380009, India}
% \affil[2]{Dept. of Physics, Indian Institute of Technology Gandhinagar, Gujarat, 382355, India}
% \affil[3]{Center for Quantum Technologies, National University of Singapore, Singapore 117543}
% % \email{vardaan@prl.res.in}
\maketitle
\section{Introduction}
Random numbers are a pre-requisite for many cryptographic applications as proposed by Claude Shannon in 1948 \cite{shannon1948mathematical}. This requirement is typically fulfilled by use of algorithmically generated random bit stream, also referred to as pseudo random number generators (PRNGs). PRNGs have many applications ranging from digital signatures to IoT applications \cite{kietzmann2021guideline}. On the other front, military grade applications require highly secure random number generators. Quality of randomness in PRNGs relies on their computational complexity and their length. \cite{deng2017developments}. However, they can turn obsolete in cryptographic applications pertaining to advance algorithms which can dig up the low computational resolution making the apparently random bit streams, non-random. Quantum random number generators (QRNGs) provide an alternative based on their quantum unpredictability (arising from laws of quantum mechanics) \cite{jennewein2000fast}.  Many types of QRNGs have been studied both in discrete \cite{jian2011two} and continuous \cite{shen2010practical} degrees of freedom for their use in cryptographic applications. 

On the testing front, NIST Statistical Test Suite, Dieharder \cite{novark2010dieharder}, ENT \cite{walker2008ent} have been used to prove randomness of QRNGs. But this is half-baked truth. NIST Statistical Test Suite \cite{bassham2010sp}, as the name suggests, are used for testing statistical properties like uniformity amongst PRNGs and thus, can strictly be used to check for statistical properties of QRNGs.  To check for the unpredictability of PRNGs, NIST has provided a test suite based on entropy model \cite{turan2018nist}. Nonetheless for the quantum case, one has to use some quantum correlation quantifier like quantum entanglement measures \cite{Terhal_20}. For example, given n bits, these quantum correlations forbid one to predict (n+1)th bit. There is no common consensus on using a single quantum certification for all quantum random number generators and hence, different works in the literature use different techniques for certification \cite{Entropy_Mazzuchi}. By certification, we understand that a given quantifier can certify that the bits generated have a quantum correlation. This work uses quantum measures like CHSH inequality violation \cite{dehlinger2002entangled} as a quantifier.

On the device independence front, treating the process of generation and measurement as a black box requires a lot of independent checks and thus, resources. Our technique, being device-independent, assumes no constraints imposed by the devices. Such a technique can easily be integrated into faster device-independent protocols \cite{bhavsar2023improved, mahadev2022efficient} by using an efficient non-linear crystal.

The article is subdivided into following four sections. Section II discusses about the theoretical framework required to understand the physical process of generation and certification of random number generation. Section III describes the experimental setup on how to generate random numbers from bipartite quantum entanglement. Section IV discusses the random bits generated in a device independent manner. It discusses their performance against standard tests. Section V draws conclusions on different aspects of this study.

\section{Theoretical Aspects}
 Before discussing the experimental setup, we need to understand the theoretical framework which circumscribes the quantum random number generation and testing process. First, we discuss the generation process of QRNG and discuss it's source-independence. Next, we discuss the typical method (NIST Statistical Test Suite) of checking uniformity of QRNGs. Thirdly, we describe the quantifiers from quantum mechanics for certification of unpredictability of QRNGs.

\subsection{Generating random numbers from quantum entanglement}
The use of quantum entanglement for the generation of quantum random numbers, requires the ability to verifiably produce high quality entangled states. For this, we use a quantum eraser experimental setup derived from Hong Ou Mandel (HOM) interference. Once at the dip point of the HOM interference curve, we can generate entangled photons by making the photons distinguishable in one degree of freedom, say polarization analogous to the quantum eraser experiment as described in \cite{gerry2023introductory}. This is done by rotating one of the half waveplates by 45\degree. Theoretically, the output state of Hong Ou Mandel (represented by pt. A in Fig. \ref{fig:HOM}) is
\begin{equation*}
    \ket{\psi_{AB}}=(\ket{2H}_a\ket{0}_b+\ket{0}_a\ket{2H}_b)/\sqrt{2}
\end{equation*} If the polarisation of one of the photon is rotated by 45 $\degree$ to the horizontal, the state transforms to 
\begin{equation*}
    \ket{\psi_{AB}}=\frac{1}{2}(\ket{H}_a\ket{V}_b-\ket{V}_a\ket{H}_b)+\frac{\iota}{2}(\ket{HV}_a\ket{0}_b+\ket{0}_a\ket{HV}_b)
\end{equation*}
Because our detectors are click detectors, the post-selected state reduces to Eq. \ref{State}. Finally, we generate random numbers through a bipartite entangled state. Mathematically, a maximally entangled state in polarisation can be written as

\begin{equation}\label{State}
    |\psi_{AB}\rangle=\frac{1}{\sqrt{2}}(|H_a\rangle_{a}|V\rangle_{b} - |V\rangle_{a}|H\rangle_{b})
\end{equation}

and, 
\[p_{0}=\operatorname{Tr}\left((M_{0}\otimes I) (\rho_{AB})\right) =\frac{1}{2}\]

and,
\[p_{1}=\operatorname{Tr}\left((M_{1} \otimes I) (\rho_{AB})\right) =\frac{1}{2}\]
 
\noindent where, $\rho_{AB}$ is the bipartite density matrix of the entangled state, $M_{0}$ and $M_{1}$ are projective measurement operators \cite{nielsen_chuang_2010} and ``0" and ``1" refers to binary bits. Furthermore, by tracing out one of the qubits of a bipartite quantum entangled pair, the other qubit has an equal probability of being either H or V polarized (represented by values $p_{0}$ and $ p_{1}$) respectively. This equiprobable distribution is the basis of generating random numbers from a bipartite entangled state. We employ HOM curve as a certification technique where it's visibility serves as a measure of source independence. 

\subsection{NIST Statistical Test Suite}
The NIST Statistical Test Suite is a hypothesis based suite which checks whether a given sequence is random. The test-static \textit{p} gives the confidence level in the hypothesis with a threshold set to 0.01. The test suite entails 15 tests with numerous sub-tests. These tests check for patterns in the bit-stream in long as well as short range. The tests include auto-correlation, compression and spectral frequency based tests \cite{bassham2010sp}. In other words, they check for uniformity of the bit stream. Before testing against NIST test suite, the data is post-processed using Toeplitz hash function. The advantage of using Toeplitz hash function is that it scrambles the "quantum information" evenly amongst all bits. This is ensured by 2-universality of Toeplitz hash function \cite{ma2013postprocessing}.

\begin{figure*}[!htp]
\centering
\includegraphics[scale=0.5]{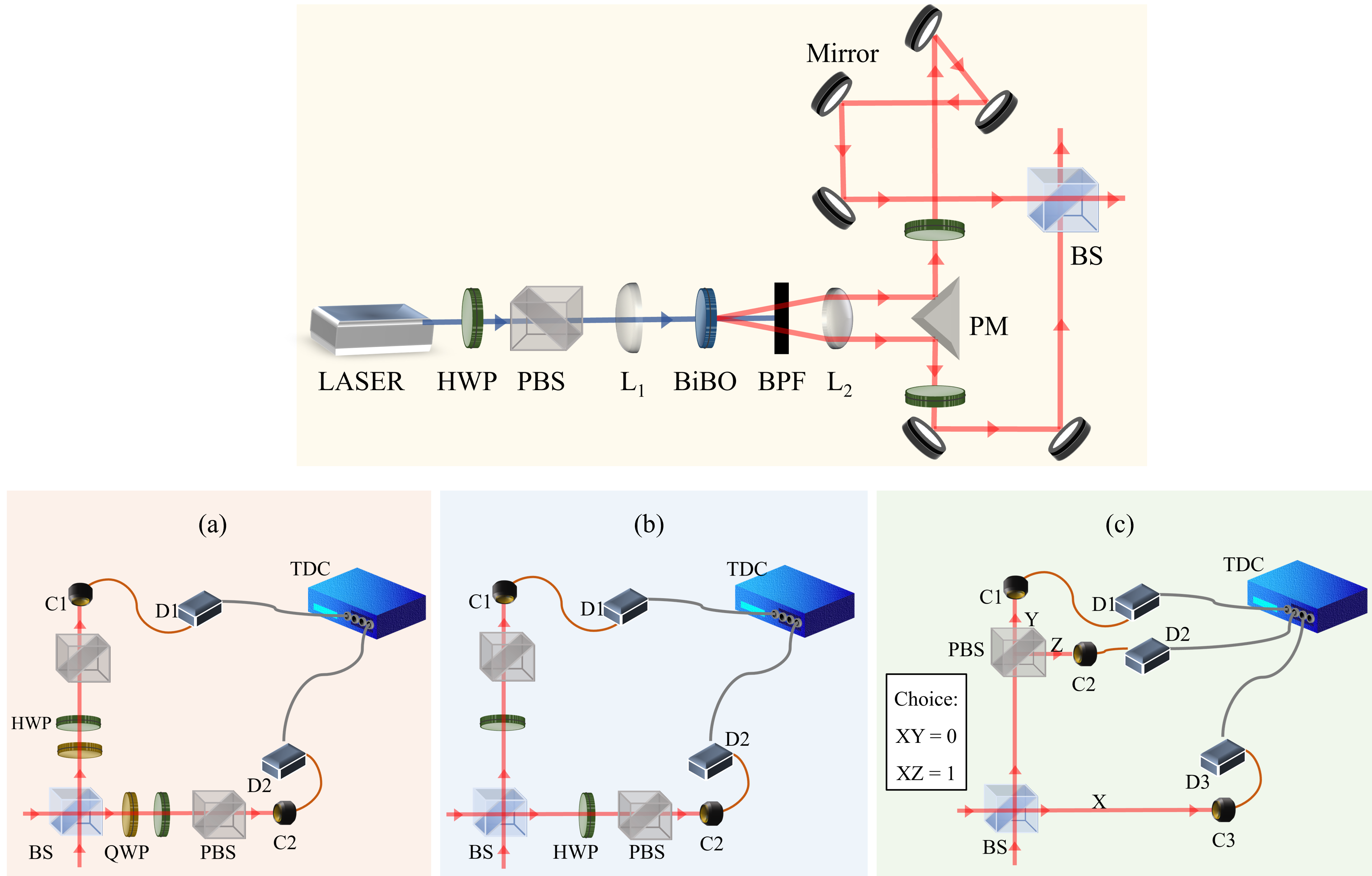}
\caption{Experimental Setup for entanglement based random numbers: (a) certification via density matrix (b) certification via direct CHSH Bell parameter (c) generation of random numbers. 
$L_{1},L_{2}$: lens, PM: prism mirror, BS: beam splitter, PBS: polarising beam splitter, QWP: quarter wave plate, HWP: half wave plate, BPF: bandpass filter, $C1, C2, C3, C4$: Coupler, D1, D2, D3: detectors, TDC: time to digital converter, ``X", ``Y" and ``Z" denote path of the photons to different detectors}
\label{fig:ExperimentalSetup}
\end{figure*}

\subsection{Theoretical framework to study Quantum Entanglement}

Several authors \cite{Wootters_98} have provided methods to quantify ``entanglement" as a resource with the aid of different measures \cite{Horodecki_1961, Horodecki_97, Terhal_20, Lewenstein_2000}. One of the well studied experimental entanglement measure is CHSH inequality \cite{CHSH1969}. The CHSH inequality measures correlations between two qubits using different measurement bases. For an entangled pair \(A\) and \(B\), we will label the bases for the qubit \(A\) to be \(a\), \(a'\), and  for qubit \(B\) to be \(b\), \(b'\) respectively.  This allows us to calculate the CHSH parameter \(S\),
\begin{equation}
\label{Eq: CHSH}
S = E(a,b) - E\left(a,b'\right) + E\left(a',b\right) + E\left(a',b'\right),\,\,
\end{equation}
\newline 
where,
E(\(a,b\)),\,\, represents the expectation value of the product of measurements in \(a\)-\(b\) basis. The same holds true for other combinations of \(a'\)-\(b'\) basis. For CHSH parameter $S\geq2$, the measured state shows quantum entanglement. 

Alternative to this measure, quantum state tomography (QST) estimates the density matrix, $\rho$ describing a given state from projective measurements. Pertaining to statistical noise in such experiments, Bayesian and Maximum likelihood methods are used in finding the estimate $\rho_{est}$ closest to true density matrix $\rho$ from the space of physical density matrices $\mathbf{P}$ \cite{schmied2016quantum}. The premise to choose CHSH inequality is that it is well studied while premise for density matrix choice is that it contains complete information about a state.

Consider a 2-qubit state characterized by its density matrix \(\rho\), which can be written as follows:
\begin{equation}
   \label{Eq: State} 
   \rho  = {\frac{1}{2^2}}\sum_{i_1, i_2 = 0}^{3} u_{i_1,\,i_2} \hat{\sigma}_{i_1}\otimes\hat{\sigma}_{i_2},
\end{equation}
where ${u_{i_1,\,i_2}}$ are real numbers and $\hat{\sigma}_{i}$ are the Pauli matrices. The state $\rho$ can be obtained from $4^2$ (for a  2-qubit system and 4 refers to Stokes parameters \cite{schaefer2007measuring} type measurements) optimal measurements \cite{Kwiat_2001}. To assign a density matrix to a physical state, (i) $\rho$ must be normalized ($\rm{Tr} \rho = 1$), (ii) Hermitian ($\rho^\dagger = \rho$), and (iii) positive semi-definite ($\langle \psi |\rho| \psi \rangle\,\,\geq 0$) for all unit-norm states $| \psi \rangle$.

On the initial investigation of the outcomes of projection of state $\rho$ with sufficient number of repetitive measurements, we employ least-squares (LS) inversion to estimate $\rho_{LS}$ of state $\rho$; from the projection probabilities approximated from measured outcome frequencies. This method is limited by contribution of statistical noise and may return the state $\rho_{LS}$ to be nonphysical ($\rho_{LS} \notin \mathbf{P}$), as it cannot guarantee positive semi-definiteness.

Nevertheless, one can implement the approach of estimating $\rho_{est}$ from set $\mathbf{S}$ using a point estimator such as Maximum likelihood estimator\,(MLE) \cite{Kwiat_2001} or an uncertainty region estimator like Bayesian QST \cite{Lukens_2020}. The need to implement two different approaches of estimation is two fold. One, it keeps a check on statistical noise i.e. in a device independent setting. Secondly, it draws a comparison between two different approaches of estimation. 

\subsubsection*{Bayesian quantum state estimation}
% Bayesian QST, explicitly incorporates experimental uncertainty through Bayes’ theorem. 
Consider that any state ($\rho'$) in $\mathbf{S}$ is parameterized by a vector $\rm{x}$, ensuring that any value within $\rm{x}$’s support yields a physical matrix. According to Bayes’ theorem, the posterior probability distribution of $\rm{x}$, given experiment results $D$ \cite{Lukens_2020}, follows:
\begin{equation}
    \pi(\rm{x}) = \frac{1}{Z} L_D(\rm{x})\pi_0(\rm{x}),
\end{equation}
where $L_D(\rm{x})$ represents the probability of obtaining the observed outcomes given state $\rho_{LS}$, $\pi_0(\rm{x})$ denotes the prior distribution (reflecting beliefs about $\rho$ before the experiment), and $Z$ is a normalizing constant ensuring $\int d\rm{x}\, \pi(\rm{x}) = 1$. 

Access to $\pi(\rm{x})$ enables computation of the expectation value of any function $\phi$ of $\rho'$:
\begin{equation}\label{Eq: Integral}
\langle \phi(\rho') \rangle = \int d\rm{x} \, \pi(\rm{x})\, \phi(\rho'(\rm{x})),    
\end{equation}
facilitating determination of the mean and standard deviation of any quantity of interest.

However, the numerical computation of integral resembling in Eq.\,\eqref{Eq: Integral} proves to be quite challenging. Hence, we have used method of obtaining $R$ samples $\{x^{(1)}, x^{(2)}, \ldots , x^{(R)}\}$,to approximate the Eq.\,\eqref{Eq: Integral} to:
\begin{equation}
\langle{\phi}(\rho')\rangle \approx \frac{1}{R} \sum_{r=1}^{R} \phi(\rho'(x^{(r)})),
\label{Eqn: Bayesian}
\end{equation}
 as described in \cite{Lukens_2020}. Lukens parameterise the weights from Gamma distributed random variables for the prior for $\pi_0(\rm{x})$ and use Monte-Carlo sampling methods for sampling of R which together influence the outcome $\langle{\phi}(\rho')\rangle$. As sampling (performed via help of random numbers) is an integral component in estimation of density matrix, density matrix calculations form a feedback loop with the process of random number generation. Thus, to avoid detection side attacks in a device independent setting and also draw a comparison, another estimator like maximum likelihood is required. 

\subsubsection*{Maximum likelihood quantum state estimation}

Maximum likelihood estimation (MLE) identifies the density matrix most likely to have produced the observed data $D$:
\begin{equation}
\rho_{MLE} = \text{argmax}_{\mathbf{P}}\,\,\,(L_D(\rho')),\,\, \rm{where}\,\, \rho' \in \mathbf{P}
\label{Eqn: MLE}    
\end{equation}

\noindent where $L_D \propto P(D|\rho')$ represents the probability of obtaining the observed outcomes $D$ given state $\rho'$, as defined by a specific model \cite{Kwiat_2001}. Proper parameterization of $\rho'$ ensures that the estimate ($\rho_{MLE}$) is a physical state. This method has become the prevailing approach to Quantum State Tomography (QST) due to its easiness. However, $\rho_{MLE}$ is a point estimate, lacking quantification of result uncertainty. 

Once the density matrix is calculated, we evaluate CHSH Bell parameter S, as an upper bound, given by the formula, 
\begin{equation}
S_{max}=2\sqrt{c_{11}^2+c_{22}^2}
\label{Eq: CHSH_RHO}
\end{equation}
where $c_{11}$ and $c_{22}$ represent the largest eigenvalues of $C^T$C where $C^T$ is the transpose of C. The correlation matrix C with elements $c_{ij}$ is related to the density matrix $\rho$ for a 2-qubit system by the relation

\begin{equation}
\rho=\frac{1}{4}\sum_{i,j=0}^{3} c_{ij}\hat\sigma^1_{i}\otimes\hat\sigma^2_{j}
\end{equation} 

as mentioned in \cite{horodecki1995violating}.

\begin{figure}[!h]
\centering
\includegraphics[scale=0.27]{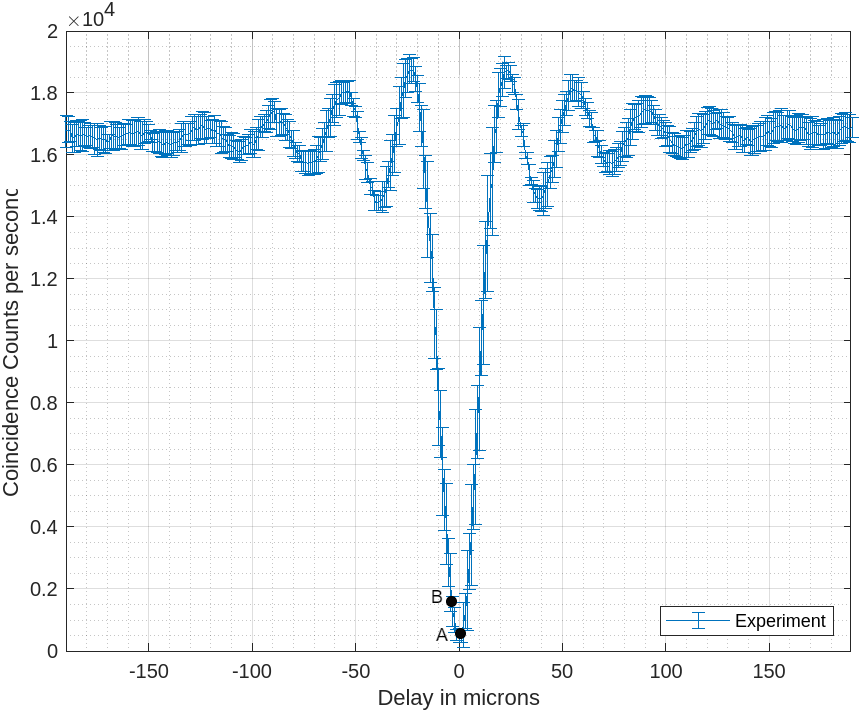}
\caption{Hong Ou Mandel (HOM) curve serves as a source-independence certification for a polarisation entangled state at points A and B}
\label{fig:HOM}
\end{figure}

\section{Experimental Setup}
The laser (TOPTICA-TopMode 405) with a center wavelength of 405 nm and bandwidth of 0.01 pm emits a Gaussian beam. The beam passes through a half-wave plate (HWP) followed by a polarising beam splitter (PBS). This combination gives a control over the intensity of the beam. A 50cm lens ($L_1$) is used to focus the beam on a Type-I Bismuth Borate (BiBO) crystal of length 5 mm. The pump power before the nonlinear crystal is 5 mW. Two correlated degenerate spontaneous parametric down-converted photons are generated in the crystal in a non-collinear geometry assisted via angle tuning. The photon pairs are collimated by a 10 cm lens ($L_2$) and separated by a prism mirror (PM) to interfere in a Mach Zehnder like setup as shown in Fig. \ref{fig:ExperimentalSetup}. In one of the paths, a motorized translation stage MTS25-Z8, with a resolution of 29 nm, is added to compensate for the extra delay, if any. The two output ports of the beam splitter are initially used to show a HOM dip \cite{HOM} as shown in Fig. \ref{fig:HOM}.

Once the two polarisation entangled qubits are generated as described in the theoretical section, one can simply detect one of the qubits in the detector following path ``X". The other qubit is measured in H-V polarisation basis (path ``Y" and ``Z"). Measuring this other qubit in coincidence with the photon in path ``X" generates random numbers as it gets detected either in the H polarisation (bit 0) or V polarisation (bit 1) at a particular instant with a 50:50 probability as shown in Fig. \ref{fig:ExperimentalSetup} (c) .The detectors (SPCM-800-14FC with a dark count of 100 counts per second) are identical in all three arms.
To see measurement-device independence on the detector side, the diagram of the setup is slightly modified after the beam splitter to take projective measurements using quarter wave plate (QWP), HWP and a PBS combination as shown in Fig. \ref{fig:ExperimentalSetup} (a). On the contrary, to show the source independence, HOM curve is used as a certification technique. it's visibility serves as a parameter for source-independence. Thus, a high visibility of the HOM curve is desired. In our case, we achieved a visibility of 97\%. Since no phase information is required to generate bit streams, random bit-streams generated from maximally mixed state of a bipartite system and single photon source are identical. The extra phase information in the entangled case helps in providing security against attacks on the source. After calibration of data points A and B (as highlighted in Fig. \ref{fig:HOM}) using the HOM curve, we generate entangled states at these data points.

\begin{table*}[htbp]
\centering
\begin{tabularx}{0.575\textwidth}{|c|c|c|}
\hline
\textbf{NIST Statistical Test Suite} & \textbf{ \textit{p}-value(Dataset A)} & \textbf{\textit{p}-value(Dataset B)} \\
\hline\hline

Approximate Entropy & 0.985 & 0.546 \\
\hline
Block Frequency & 0.380 & 0.129 \\
\hline 
Cumulative Sums & 0.973 & 0.557 \\
\hline
FFT & 0.979 & 0.973 \\
\hline
Frequency & 0.840 & 0.465 \\
\hline
Linear Complexity & 0.840 & 0.965 \\
\hline
Longest Runs & 0.060 & 0.966 \\
\hline
Non Overlapping Template Matching & 0.069 & 0.325 \\
\hline
Overlapping Template Matching & 0.721 & 0.590 \\
\hline
Random Excursions & 0.843 & 0.383 \\
\hline
Random Excursions Variant & 0.435 & 0.621\\
\hline
Rank & 0.993 & 0.084 \\
\hline
Run's & 0.858 & 0.325 \\
\hline
Serial & 0.403 & 0.356 \\
\hline
Universal & 0.285 & 0.210 \\
\hline
\end{tabularx}
\caption{\textit{p}-value for different test suites: Dataset A corresponds to entangled state generated at Setting A and Dataset B corresponds to entangled state generated at Setting B respectively as highlighted in Figure \ref{fig:HOM}. Both post-processed datasets are 1.2 M long. }
\label{tab:uniformity}
\end{table*}

\section{Results and discussion}
In this study, we have generated two datasets of 4.5 million (M) bit streams from the experimental setup shown in Fig. \ref{fig:ExperimentalSetup}. The data is recorded for two different cases: Dataset A is recorded at the point of highest visibility of the HOM curve of Fig. \ref{fig:HOM}), which represents a maximally entangled state, and Dataset B is recorded 700 nm away from dip point which degrades the quality of the above entangled quantum state. After post-processing using Toeplitz hash function, the length of random numbers is reduced from 4.5 M to 1.2 M.

\begin{figure}[H]
     \centering
     \begin{subfigure}[t]{0.5\textwidth}
         \centering
         \includegraphics[width=0.6\textwidth]{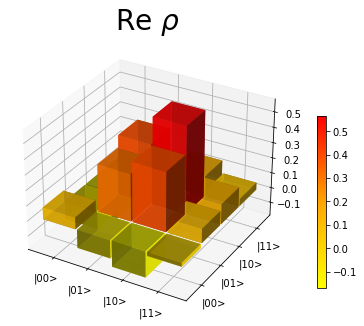}
         \caption{Real part of the density matrix}
         \label{fig1}
     \end{subfigure}
     \hfill
     \begin{subfigure}[t]{0.5\textwidth}
         \centering
         \includegraphics[width=0.6\textwidth]{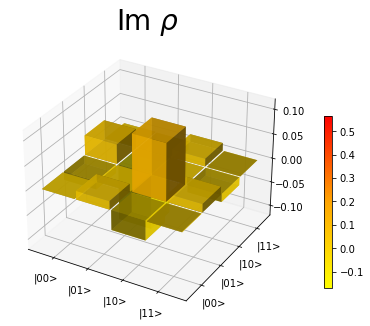}
         \caption{Imaginary part of the density matrix}
         \label{fig2}
     \end{subfigure}
    
        \caption{Projection of an entangled Bell state (Eq. \ref{State}) in H-V basis. The magnitude of each projection is color-coded. (a) and (b) indicate real and imaginary part of the density matrix at setting A of the HOM curve.}
        \label{fig:three graphs}
\end{figure}

The datasets are tested against NIST-STS for quality of statistical randomness, and the results are highlighted in Table \ref{tab:uniformity}. Kolmogorov-Smirnov test (KS test) is performed on sets with multiple p-values to provide an overall p-value indicating that they are normally distributed. Every test in Table \ref{tab:uniformity}, shows a test-static $p\,\geq\,0.01$, suggesting that the generated dataset is statistically random.

The dataset A and B are obtained from two different quantum states, therefore having different ``quantum information". Here, we have used CHSH inequality\,(S) as one such quantifier of entanglement, as given in Eq.\,\eqref{Eq: CHSH}. The CHSH violation is measured from the direct observation as discussed in Ref. \cite{dehlinger2002entangled} and alternatively by estimating the density matrix using Bayesian and MLE quantum state tomographic techniques as in Ref.\cite{Kwiat_2001, Lukens_2020}.

% For e.g.,\, the reconstructed density matrix $\rho_{MLE}$ of projected entangled Bell state given in Eq.\eqref{Eq: State} using an MLE estimator is plotted in Fig.\,\ref{fig:dmd}. Note that the real and imaginary parts of $\rho_{MLE}$ are shown separately with the color coding of its magnitude.
 For the estimation of S, the direct experimental measurements are taken using the method outlined in Ref.\,\cite{dehlinger2002entangled}, and results are shown in column 2 of Table\,\ref{tab:privacy}. The S for the settings in dataset A and dataset B are well above the classical limit $(|S_{classical}|=2)$, indicating the quantum nature of random bits in both datasets. However, to eliminate the possibility of assumptions based on device imperfections (say, detector efficiency), we calculate $\mathbf{P}$ using Eq.\,\eqref{Eq: CHSH_RHO}. The estimated density matrices ($\rho^A_{est},\,\,\rho^B_{est}$) are the density matrices of the entangled photon states used to generate the datasets A and B respectively.

% \begin{figure}[htbp]
     %\centering
     %\includegraphics[scale=0.25]{Density_Matrix_Data.png}
     %\caption{Setting B:Figure shows the projection of an entangled Bell %state of \eqref{State} in H-V basis. The magnitude of each %projection is color-coded.}
   %  \label{fig:dmd-B}
% \end{figure}

 To estimate the density matrix for each dataset using MLE, experimental measurements were performed as in Ref.\,\cite{Kwiat_2001}. Column 3 of Table\,\ref{tab:privacy} provides the estimated value of {S($\rho_{MLE})$} using Eq.\,\eqref{Eq: CHSH_RHO}. For the settings in dataset A , we estimate ${S(\rho^A_{MLE})}\,=\,2.65$ and for dataset B, ${S(\rho^B_{MLE})}\,=\,2.40$. The estimated $\rho^A_{MLE}$ is shown in Fig.\,\ref{fig:three graphs}.
\begin{table}[]
    \resizebox{0.48\textwidth}{!}
    % \resizebox{\columnwidth}{!}
    {%
    \begin{tabular}{| c | c | c | c |}
    \hline
     Sr. No.     & S & S($ \rho_{MLE} $) & S($ \rho_{Bayesian} $)\\
    \hline
    Dataset \hspace{0.05cm} A  & 2.78 $\pm$ 0.03 & 2.65 & 2.81 $\pm$ 0.02   \\
    \hline 
    Dataset \hspace{0.05cm} B  & 2.51 $\pm$ 0.02 & 2.40 & 2.47 $\pm$ 0.01   \\ 
    \hline
    \end{tabular}%
    }
    \caption{CHSH Bell parameter $S$ value obtained via different methods is shown. Column 2 shows direct S measurement; Column 3 denotes S obtained from $\rho$ post-processed with MLE; Column 4 denotes S obtained from $\rho$ post-processed with Bayesian estimation. $|S|\geq2$ indicates quantum behaviour of the device.}
    \label{tab:privacy}
\end{table}

The results in Table\,\ref{tab:uniformity} indicate a high confidence that both the datasets have statistically independent random bit-streams. Additionally, the results shown in Table\,\ref{tab:privacy}, prove the quantum correlation. The table, although indicating quantum signature, shows some inconsistency amongst different values. This discrepancy arises due to two reasons. One that Bayesian estimation is better than MLE in cases where right prior is known (Dataset A in our case). Secondly, the effect of sampling can bias the Bayesian estimation (Dataset B in our case). For Dataset B, right prior is not known. This requires further investigation. However, the result provided in this study positively concludes that the QRNG developed using entangled polarisation entangled photon pair is statistically random and secure against source and detector side attacks as indicated by HOM dip visibility and CHSH value respectively. A contrasting difference between our technique and randomness expansion protocols is that we have performed sequential measurements bunched together for a quantum correlation quantifier, while the semi-device independent protocols sparse them in generation and test rounds with some bias picked from a quantum source.

One open problem with random numbers is to investigate whether unpredictability and statistical properties are inter-related. To address this query, we performed scrambling operations at different lengths. One can use a quantifier from information theory, specifically min-entropy \cite{konig2009operational}, defined as:
 \begin{equation*}
      H_{\infty} (X)= - log_2 (p_{max})    
 \end{equation*}
where $p_{max}$ is the probability of maximum occurrence of random variable X to quantify the said relationship.
It is observed that min-entropy for Dataset A is $H_{\infty}(X)=0.999735$ and for Dataset B is $H_{\infty}(X)=0.999038$. This decrease shows a relation between quantum unpredictability and statistical properties. As there is a decrease in statistical correlations (refer to p-values in Datasets A and B of the frequency test), there is a similar decrease in quantum unpredictability. One possible explanation is that the scrambling process is classical, and thus, it is challenging to see quantum unpredictability signature in highly classically processed data. This is supported by the fact that the p-values of the Frequency test and Linear Complexity test are equal for Dataset A. These observations require further investigation of the parameters of the algorithms used in the test suite.

\section{Conclusion}
In this article, we have provided a sanity check of the quantum source (HOM dip) and investigated the working of a fully device-independent quantum random number generator. One can put a constraint on resources and translate the fully device-independent scheme to measurement device-independent or source-independent quantum random number generator protocols with automation of HWPs and calculation of smooth conditional min-entropy (based on density matrix). Other quantifiers like entanglement measure/witness can be used to define new protocols in semi-device independent regime or for higher dimensions. Also, it is theoretically proven that the visibility of the HOM curve is equal to the purity of input photons \cite{branczyk2017hong}. This provides a correlation between closeness to the dip point and the amount of CHSH Bell violation as indicated by Table \ref{tab:privacy}. Physically, it means that relative phase information between two photons is used to prove device independence and thus, preventing attacks on QRNGs from Eve. Here device independence checks can be relaxed to cases where HOM serves as a check for source independence or CHSH Bell parameter serves as a security check for measurement device independence. Sending the qubit which is being traced out here, to another party, post-processing to consider loss over the channel and together with the addition of local operations using classical communication (LOCC) can convert this random number generation scheme to device-independent quantum key distribution scheme.

\bibliographystyle{unsrt}
\bibliography{main}

\end{document}